\documentclass[10pt]{article}

\usepackage{amssymb}
\usepackage{epsfig}

\begin{document}

\title{Is There a Hollow Inside the Proton?\footnote{A talk at the XXXI-th International Workshop on High Energy Physics 
``Critical points in the modern particle physics'', July 5-7, 2017, Protvino, Moscow region, Russia.}}

\author{V. A. Petrov and A. P. Samokhin \\
\textit{A.A. Logunov Institute for High Energy Physics}\\
\textit{of NRC ``Kurchatov Institute''}\\
\textit{Protvino, 142281, Russian Federation}}

\date{}

\maketitle

\begin{abstract}
We discuss a recently proposed interpretation of some model descriptions of the proton-proton elastic scattering data as a manifestation 
of alleged relative transparency of the central part of the interaction region in the impact parameter space. We argue that the presence of
nonzero real part of the elastic scattering amplitude in the unitarity condition enables to conserve the traditional interpretation.
\end{abstract}


\section{Introduction}
One of the striking properties of the elastic pp scattering is a growth with energy of the value of the ratio of the elastic to total cross-section 
$ \sigma_{\mathrm{el}}(s)/\sigma_{\mathrm{tot}}(s)$ [1, 2]. This growth was treated as an increase of the interaction intensity in the central part
of the interaction region (which grows with energy) in the impact parameter space [1]. However, in connection with the LHC data [2], which reveal 
that the ratio $ \sigma_{\mathrm{el}}(s)/\sigma_{\mathrm{tot}}(s) $ overcomes the ``critical'' value 0.25, some authors saw a crisis in such an 
interpretation. 
More specifically, it is about the ``critical'' value of imaginary part $ a(s,b)$ of the elastic scattering amplitude in the impact parameter space
$ f(s,b)=r(s,b)+ia(s,b)$, which satisfies the $s$-channel unitarity condition
\begin{equation}
a(s,b) = a^{2}(s,b) + r^{2}(s,b) + \eta(s,b),\,\,\,\,\, \eta(s,b) \geq 0.
\end{equation}
The inelastic overlap function (or the profile function, or the inelasticity profile)
\begin{equation}
P_{\mathrm{inel}}(s,b) \equiv 4\eta(s,b) = 4[a(s,b) - a^{2}(s,b) - r^{2}(s,b)],\,\,\, 0 \leq P_{\mathrm{inel}}(s,b) < 1
\end{equation}
is the probability of an inelastic pp interaction at the energy $ \sqrt{s}$ and impact parameter $b$. If $ P_{\mathrm{inel}}(s,b) = 1$ at some
finite values of $s$ and $b$, then the elastic pp scattering is impossible at these values of $s$ and $b$. Such a scenario looking nonsensical,
we consider that $ P_{\mathrm{inel}}(s,b) < 1$ at any finite values of $s$ and $b$.

Since the real part $ r(s,b)$ of the elastic scattering amplitude is much less than imaginary part $ a(s,b)$ another function, 
\begin{equation}
G_{\mathrm{inel}}(s,b) \equiv 4a(s,b)[1-a(s,b)]
\end{equation}
is analysed often instead of $ P_{\mathrm{inel}}(s,b)$.
We must remember that only $ P_{\mathrm{inel}}(s,b)$ has the above physical meaning while $G_{\mathrm{inel}}(s,b)$ is only 
$ \approx P_{\mathrm{inel}}(s,b)$. It is easy to see that the signs of the partial derivatives of $G_{\mathrm{inel}}(s,b)$ and $ a(s,b)$
\begin{equation}
\frac{\partial G_{\mathrm{inel}}}{\partial s} = 4(1-2a)\frac{\partial a}{\partial s}\,,\,\,\,\,\, 
\frac{\partial G_{\mathrm{inel}}}{\partial b} = 4(1-2a)\frac{\partial a}{\partial b}
\end{equation}
are the same if $ a(s,b) < 1/2 $, but these are opposite if $ a(s,b) > 1/2 $.

Let us assume (in accordance with the analysis of the experimental data) that the imaginary part of the scattering amplitude $ a(s,b)$ is 
a monotonically decreasing function of $b$ (at fixed value of $s$) and a monotonically increasing function of $s$ (at fixed value of $b$),
i.e. $ \partial a(s,b)/\partial s > 0$ and $ \partial a(s,b)/\partial b < 0$.
Then according to Eq. (4), until $a_{0}(s) \equiv a(s,b=0) < 1/2 $, the behaviour of function $ G_{\mathrm{inel}}(s,b)$ is similar to that 
of $ a(s,b)$ (see Fig. 1). But at very high energies, when $a_{0}(s)$ overcomes the ``critical'' value 1/2, the further growth of $ a(s,b)$
leads to decreasing of $ G_{\mathrm{inel}}(s,b)$ with energy at small impact parameters (i.e. in the ``head-on collisions'' region) 
and to formation of the peripheral profile (see Fig. 2).
Such paradoxical behaviour of the approximate inelasticity profile $ G_{\mathrm{inel}}(s,b)$ was discussed in Refs. [3--13] as a new feature
of the high energy hadron-hadron interaction (for example, as a ``hollowness effect'' --- minimum of the inelasticity profile at zero impact 
parameter [13]).
\begin{figure}
\begin{center}
  \parbox{2.1in}{\includegraphics[width=2in]{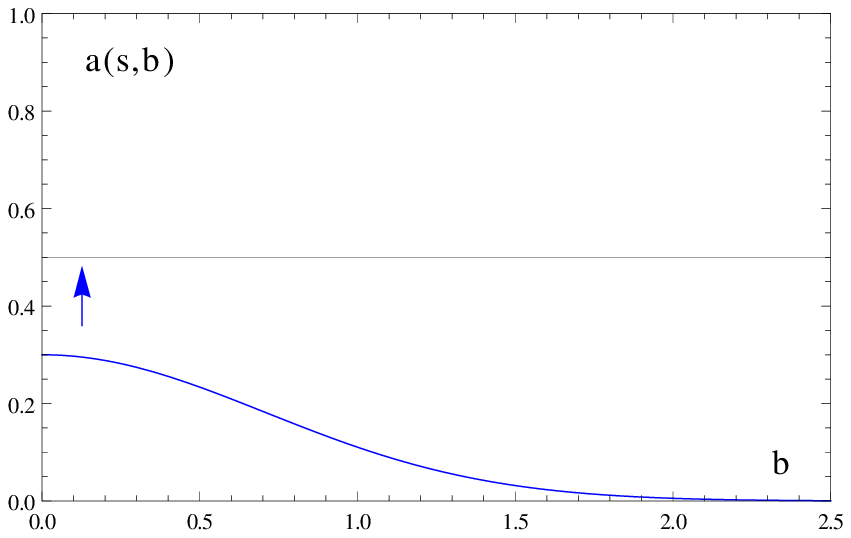}}
  \hspace*{4pt}
  \parbox{2.1in}{\includegraphics[width=2in]{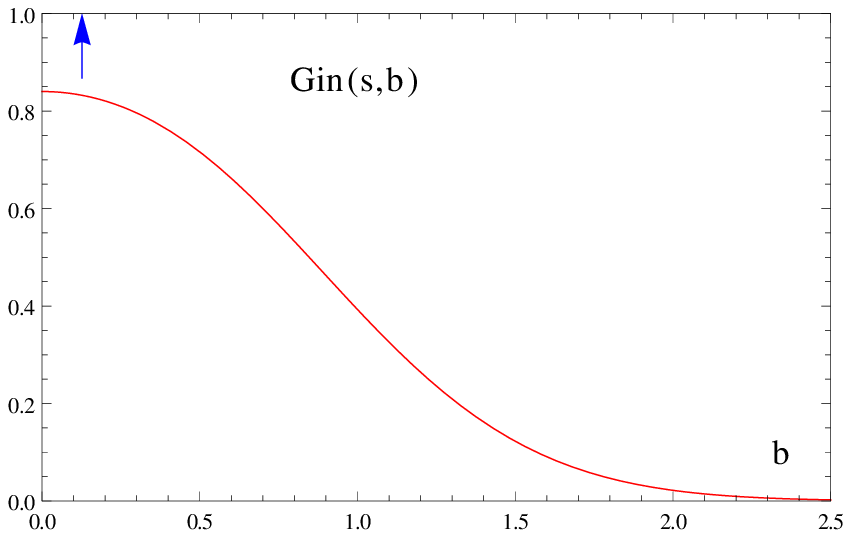}}
  \caption{When $a_{0}(s) \equiv a(s,b=0) < 1/2 $ the $ G_{\mathrm{inel}}(s,b)$-function (the right panel) behaves itself similar 
  to $ a(s,b)$ (the left panel). The profiles of $ a(s,b)$ and $ G_{\mathrm{inel}}(s,b)$ grow with energy as the arrows indicate.}
\end{center}
\end{figure}
\begin{figure}
\begin{center}
  \parbox{2.1in}{\includegraphics[width=2in]{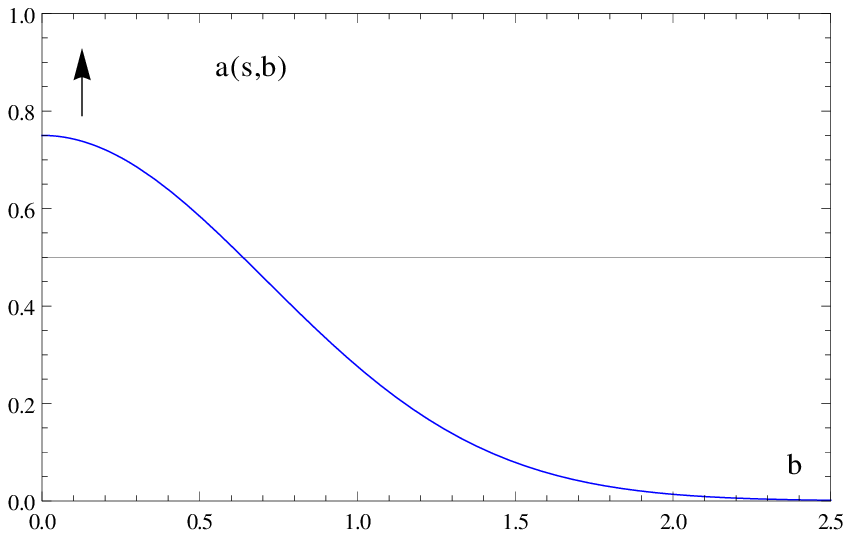}}
  \hspace*{4pt}
  \parbox{2.1in}{\includegraphics[width=2in]{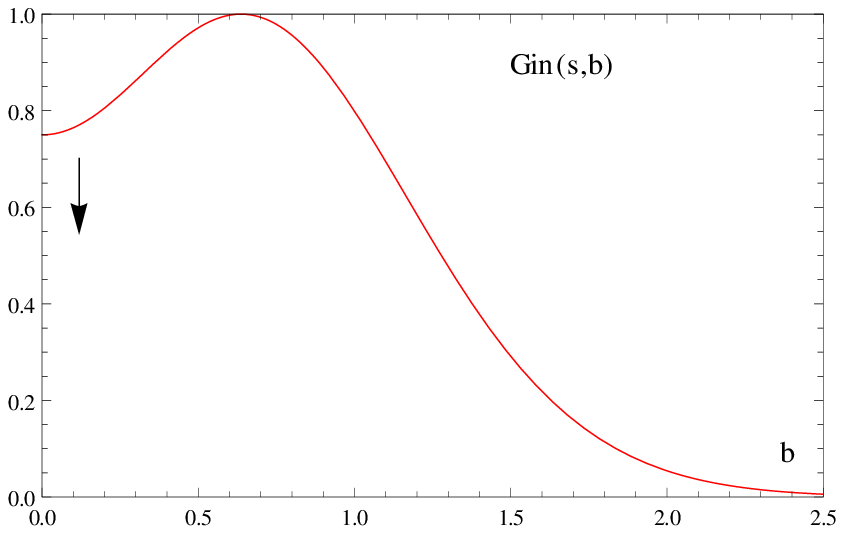}}
  \caption{When $a_{0}(s) > 1/2 $ and $ a(s,b)$-profile (the left panel) grows with energy the $ G_{\mathrm{inel}}(s,b)$-function 
  (the right panel) decreases with energy in the central region and develops a peripheral profile.}
\end{center}
\end{figure}
\section{The role of a real part of the scattering amplitude}
As the functions $ a(s,b)$, $ r(s,b)$, $ \eta(s,b)$ are not observables their reconstruction from experimental data has a model-dependent character.
The simplest model --- a pure imaginary scattering amplitude $\sim exp(Bt/2)$ --- gives [5]
\begin{equation}
a(s,b=0) \equiv a_{0}(s) = 2 \sigma_{\mathrm{el}}(s)/\sigma_{\mathrm{tot}}(s).
\end{equation}
Since at the LHC energies $\sigma_{\mathrm{el}}(s)/\sigma_{\mathrm{tot}}(s) \gtrsim 0.25 $ [2], this model leads to $a_{0}(s) > 1/2 $
in this energy region. Treatments of the 7 TeV TOTEM data (based on the amplitude parametrization [14])
give both $a_{0}(s) > 1/2 $ [3] and $a_{0}(s) \approx 1/2 $ [15]. Nevertheless, the same data are fairly well described in a Regge-eikonal model [16], 
where $a(s,b) < 1/2 $. So, it is still too early to draw any definite conclusions about the value of $a_{0}(s)$ at the LHC energies [4, 5].

Yet, let us suppose that at the LHC energies the value of $a_{0}(s)$ overcomes 1/2 and continues to grow with energy. Then
$ G_{\mathrm{inel}}(s,b)$ has a peripheral profile (see Fig. 2). Now, let us see what happens with the probability of inelastic interaction
\begin{equation}
P_{\mathrm{inel}}(s,b)=G_{\mathrm{inel}}(s,b)-4r^{2}(s,b)\,.
\end{equation}
Does it have a peripheral profile also? No, it is not nessesary. Due to the second term in Eq. (6) there is a possibility to have a normal
central behaviour of $P_{\mathrm{inel}}(s,b)$ at any fixed value of energy, i.e. $\partial P_{\mathrm{inel}}/\partial b < 0 $, $0 < b < \infty$ 
at any fixed value of energy 
and $\partial P_{\mathrm{inel}}/\partial s > 0 $, $s > s_{0}$ at any fixed value of the impact parameter. It is not difficult to construct 
a wide class of realistic amplitudes, which give such properties of $P_{\mathrm{inel}}(s,b)$.

Let us use the standard parametrization of the elastic scattering amplitude
\begin{equation}
a(s,b)=0.5[1-(\cos2\chi)\exp(-2\Omega)]\,,\,\,\,\,\, r(s,b)=0.5(\sin2\chi)\exp(-2\Omega)\,,
\end{equation}
\begin{equation}
\eta(s,b)=0.25[1-\exp(-4\Omega)]\,,\,\,\,\,\, \Omega(s,b) > 0\,,
\end{equation}
where $\exp(-2\Omega(s,b))$ is the inelasticity parameter and $\chi(s,b)$ is the phase shift. Expressions (7), (8) satisfy 
the $s$-channel unitarity condition (1) identically. It is well known that at any fixed energy and $b\rightarrow\infty $ the functions
$a(s,b)$, $r(s,b)$ and hence $ \eta(s,b)$ go to zero as $\exp(-b/b_{0})$, $b_{0}=const$, because the nearest to $t=0$ singularity of 
the elastic scattering amplitude $ T(s,t)$ lies at $t=4\mu^{2}>0$. It means (see Eqs. (7), (8)) that the functions $\Omega(s,b)$ and $\chi(s,b)$
also have the asymptotic behaviour $\exp(-b/b_{0})$ at $b\rightarrow\infty$. So, we can suggest that
\begin{equation}
\Omega(s,b)=\Omega_{0}(s)E(s,b),\,\chi(s,b)=\chi_{0}(s)E(s,b),\,E(s,b)=\exp[R-\sqrt{R^{2}+b^{2}/b_{0}^{2}}],
\end{equation}
where $R(s)$ is some monotonically increasing function of energy, $R_{0}<R(s)<\infty$. The function $E(s,b)$ monotonically increases with energy
(at any fixed value of the impact parameter) and monotonically decreases with growth of the impact parameter (at any fixed value of energy).
Let us assume that $\Omega_{0}(s)$ and $\chi_{0}(s)$ in Eq. (9) are also some monotonically increasing functions of energy, i.e.
\begin{equation}
d\Omega_{0}(s)/ds > 0,\,\,\,0\leq\Omega_{0}(s)<\infty;\,\,\,\,\,d\chi_{0}(s)/ds > 0,\,\,\,-\pi/2<\chi_{0}(s)<\pi/2.
\end{equation}
Then, according to Eqs. (9), (10) the function $\Omega(s,b)$ is a monotonically increasing function of energy (at any fixed value of the impact
parameter) and a monotonically decreasing function of the impact parameter (at any fixed value of energy). Therefore, the function
\begin{equation}
P_{\mathrm{inel}}(s,b)\equiv4\eta(s,b)=[1-\exp(-4\Omega(s,b))]
\end{equation}
has exactly the same properties, i.e. $\partial P_{\mathrm{inel}}/\partial b < 0$, $0 < b < \infty$ at any fixed value of energy 
and $\partial P_{\mathrm{inel}}/\partial s > 0$, $s > s_{0}$ at any fixed value of the impact parameter. So, the above monotony of $\Omega(s,b)$
ensures a normal central profile of $P_{\mathrm{inel}}(s,b)$ at any fixed value of energy. The imaginary part of the scattering amplitude 
$a(s,b)$ also has a central profile at any fixed value of energy (up to an extremely high energy) but the real part of the scattering amplitude, 
$r(s,b)$, has at high energies 
a peripheral profile (see Eqs. (7), (9), (10)). Peripherality of the $ G_{\mathrm{inel}}(s,b)$ and that of $4r^{2}(s,b)$ cancel each other 
in Eq. (6) to give a central profile of $P_{\mathrm{inel}}(s,b)$.

Now, the detailed properties of $a(s,b)$ and $r(s,b)$ are determined by functions $\chi(s,b)$ and $\Omega(s,b)$. Due to Eqs. (7), (8)
the unitarity relation (1) can be written as
\begin{equation}
r^{2}(s,b)+(0.5-a(s,b))^{2}=[0.5\exp(-2\Omega(s,b))]^{2}\,,
\end{equation}
i.e. the values of functions $r(s,b)$, $a(s,b)$ lie on the circle of radius $0.5*\exp(-2\Omega(s,b))$ with the centre in the point
($r=0$, $a=0.5$), and a position of the point ($r(s,b)$, $a(s,b)$) on this circle is given by the value of phase $-\pi<2\chi(s,b)<\pi$. 
So, the evolution of an amplitude $f(s,b)=r(s,b)+ia(s,b)$ can be displayed as a trajectory in an Argand plot. 

In particular, we can display in an Argand plot the energy evolution of a $S$-wave amplitude $f_{0}(s)=r_{0}(s)+ia_{0}(s)$, where
\begin{equation}
r_{0}(s)=r(s,b=0)\,,\,\,\,a_{0}(s)=a(s,b=0)
\end{equation}
are given by Eqs. (7), (9) and (10). When the function $2\chi_{0}(s)$ grows from some negative values $\sim(-\pi)$ at low energies through
the value $2\chi_{0}(s)=0$ at the ISR energies up to $2\chi_{0}(s)\sim\pi/2$ at the LHC energies and further up to $2\chi_{0}(s)\sim\pi$
at $s\rightarrow\infty$, the point in the Argand plot moves counterclockwise as it is seen at Fig. (3). At the LHC energies $r_{0}(s)>0$,  
$a_{0}(s)>0.5$; at very high energies the function $a_{0}(s)$ reaches its maximum $ 0.5<a_{0}^{\mathrm{max}}<1$, and further 
$a_{0}(s)\rightarrow 0.5$, $r_{0}(s)\rightarrow0$. Due to Eq. (10) the value of the Argand plot radius $0.5\exp(-2\Omega_{0}(s))$ monotonically 
decreases with energy.
\begin{figure}
\begin{center}
  \parbox{2.1in}{\includegraphics[width=2in]{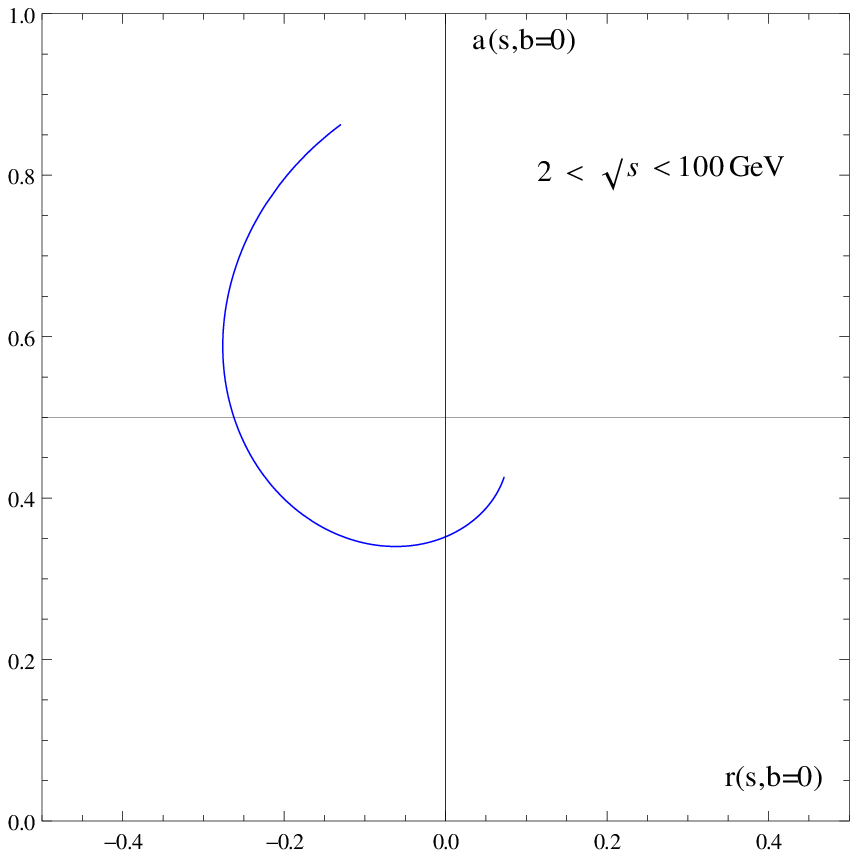}}
  \hspace*{4pt}
  \parbox{2.1in}{\includegraphics[width=2in]{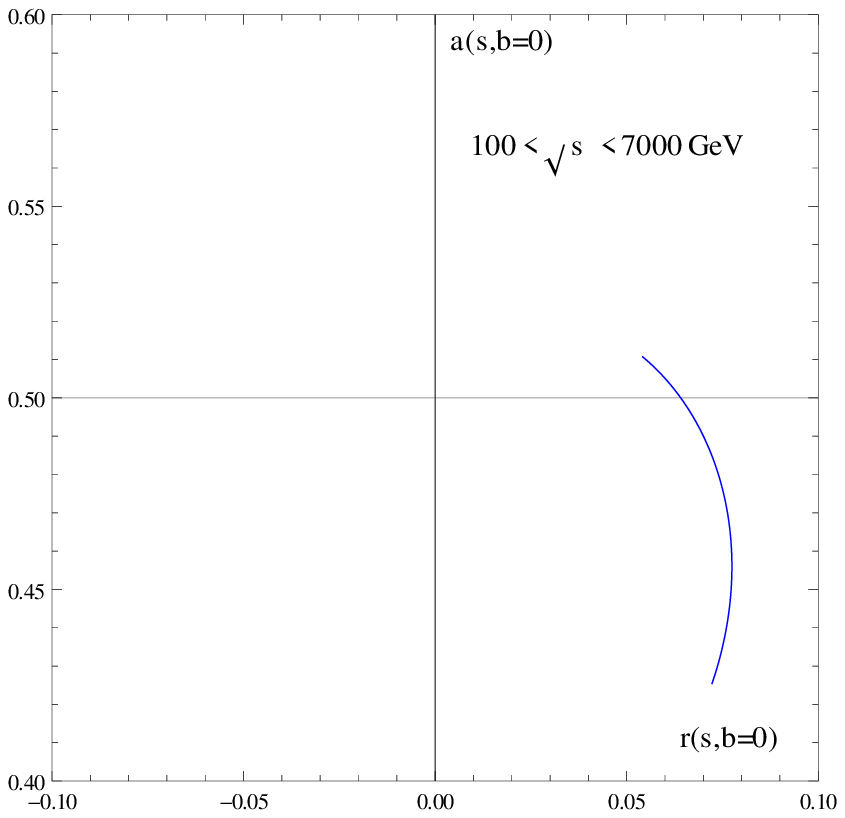}}
  \caption{Argand plot of an energy evolution of the amplitude $f(s,b=0)=r(s,b=0)+ia(s,b=0)$ at low and middle energies (the left panel) 
  and at middle and high energies (the right panel).}
\end{center}
\end{figure}
\begin{figure}
\begin{center}
  \parbox{2.1in}{\includegraphics[width=2in]{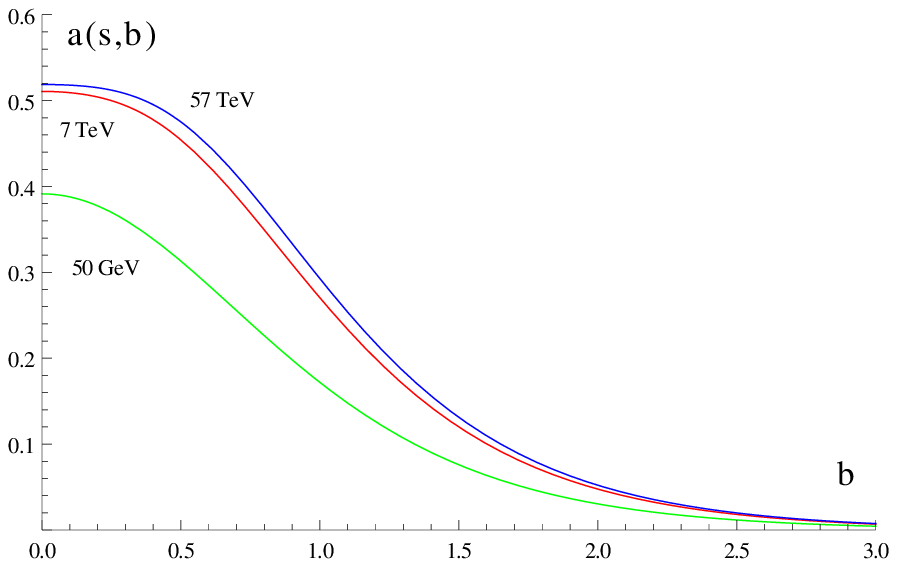}}
  \hspace*{4pt}
  \parbox{2.1in}{\includegraphics[width=2in]{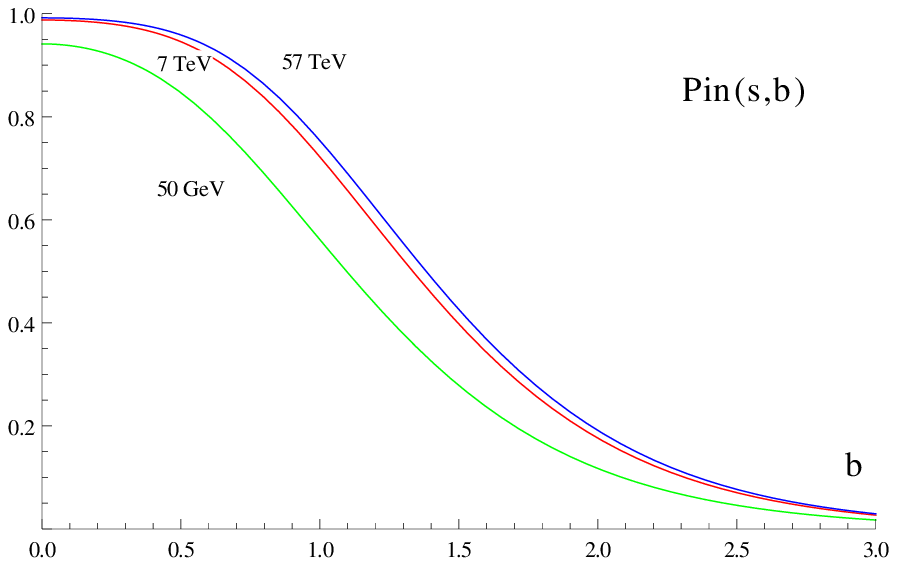}}
  \caption{When the $ a(s,b)$-profile (the left panel) grows with energy and the $a_{0}(s)$ overcomes 1/2 the $ P_{\mathrm{inel}}(s,b)$-function
  (the right panel) has a normal central profile and grows with energy also.}
\end{center}
\end{figure}

In spite of the fact that $a_{0}(s)>0.5$ at the LHC energies, the probability of an inelastic interaction $P_{\mathrm{inel}}(s,b)$ has a central 
profile at all values of energy (see Fig. 4). The central character of the $P_{\mathrm{inel}}(s,b)$ profile is ensured by the above monotony of  
$\Omega(s,b)$, but
the growth of $a_{0}(s)$ and the $r(s,b)$-evolution are related mainly with properties of the phase shift $2\chi_{0}(s)$, which is 
$\gtrsim \pi/2$ at the LHC energies
and hence $a_{0}(s) \gtrsim 0.5$, $r_{0}(s)\sim0.5\exp(-2\Omega_{0}(s))>0$. Only if the condition $r(s,b)=0$ is imposed the energy evolution 
and $b$-behaviour are given solely
by $\Omega(s,b)$ and paradoxical ``hollowness'' arises. Analysis with nontrivial $r(s,b)$ (or $\chi(s,b)\neq0$) gives more possibilities 
to understand the behaviour of the scattering amplitude.

\section{Discussion}
So, we see that the consistent account of the real part $r(s,b)$
of the scattering amplitude in the unitarity relation (1), in principle, enables to avoid the ``hollowness'' paradox. 
The inelastic overlap function $P_{\mathrm{inel}}(s,b)$ (see Eq. (2)) has a clear physical meaning, 
while the function $G_{\mathrm{inel}}(s,b)$ (see Eq. (3)) has not. At high energies, when $a_{0}(s) > 0.5$ and $a(s,b)$ has a central profile, 
peripherality of the function
$G_{\mathrm{inel}}(s,b)$ is compensated in Eq. (6) by peripherality of the $4r^{2}(s,b)$ and, as a result, the probability of an inelastic
interaction $P_{\mathrm{inel}}(s,b)$ has a central profile. Vice versa: if $P_{\mathrm{inel}}(s,b)$ has a normal central profile
at any fixed value of energy and if at high energies $a_{0}(s) > 0.5$ and $a(s,b)$ has a central profile, then the real part $r(s,b)$ of 
the scattering amplitude has a 
peripheral profile in this energy region. Peripherality of the real part of the scattering amplitude, which takes place
in the discussed model Eqs. (7)--(10), is a natural physical property of the elastic scattering in contrast with the ``hollowness''.

The impact of peripherality of the real part of the scattering amplitude on the profile function was discussed also in Ref. [17].
One of the scenarios for the real part of the scattering amplitude discussed in Ref. [13] also gives a central profile for the 
inelastic overlap function at the LHC energies.

We have discussed the qualitative properties of the model Eqs. (7)--(10), but, in our opinion, this model has a potential for a 
quantitative description of the experimental data.

To conclude, we would like to stress that a strange idea that due to some unknown cause the head-on collisions of the extended, composite objects 
should leave these objects intact is an artefact of some very specific assumptions.

\section{Acknowledgements}
We are grateful to I.M. Dremin for useful discussions.

\end{document}